\documentclass[12pt,a4paper]{article}
\usepackage{amsmath,amssymb}
\usepackage{graphicx}
\usepackage{cite}

\allowdisplaybreaks[3]

\begin{document}

\begin{flushright}
KEK-TH-2223
\end{flushright}

\begin{center}
{\Large\textbf{Holographic $\beta$ function in de Sitter space}}
\end{center}

\begin{center}
Yoshihisa \textsc{Kitazawa}$^{1),2),}$
\footnote{E-mail address: kitazawa@post.kek.jp} 
\end{center}

\begin{center}
$^{1)}$
\textit{KEK Theory Center, Tsukuba, Ibaraki 305-0801, Japan}\\
$^{2)}$
\textit{Department of Particle and Nuclear Physics}\\
\textit{The Graduate University for Advanced Studies (Sokendai)}\\
\textit{Tsukuba, Ibaraki 305-0801, Japan}\end{center}

%\date{\today}

\begin{abstract}

The scale invariance of the universe is slightly broken by slow roll parameters. 
It is likely the slow roll is dual to the random walk. We investigate the distribution function of  the conformal zeromode.
We identify de Sitter entropy $S_{dS}$ with the distribution entropy of the conformal zeromode $\rho(\omega)$. We have collected convincing support on our postulate. The semiclassical evidence is that the both are  given by the gravitational coupling $1/g=\log N/2$ where $g=G_NH^2/\pi$ and $N$ is the e-folding number. We show the renormalized distribution function obeys gravitational Fokker-Planck equation (GFP) and Langevin equations .
Under the Gaussian approximation, they boil down to a simple first order partial differential equation. 
The identical equation is derived by the thermodynamic arguments in the inflationary space-time. GFP determines the evolution of de Sitter entropy of the universe. It coincides with $\beta$ function of $g$.
We find two types of the  solutions of GFP:(1) UV complete spacetime and (2) inflationary spacetime with power potentials. 
The maximum entropy principle favors the scenario: (a) born small $\epsilon$
and (b) grow large by inflation.
We like to convey the emerging notion of de Sitter duality. The inflationary universe: (bulk/geometrical) is dual to the stochastic space-time on 
the boundary  (cosmological horizon ) as the both  are the solutions of GFP. 
%and the slow roll parameter $\epsilon \sim 1/N$.  

\end{abstract}

\newpage
\section{Introduction}

The important questions on quantum gravity are connected with their Hubble scales.
We may cite the cosmological constant problem, inflation , dark energy and so on.
It is persuasive to relate them to the infra-red behavior of quantum gravity just like hadron physics
with the infra-red behaviors of QCD. With the accumulating observational information, the time is ripe to
make a substantial progress on quantum gravity.   Consistent quantum gravity such as string theory and matrix models are manifestly UV finite. What is difficult is to understand their infra-red behavior. 

In the inflation theory, the universe undergoes the accelerated expansion. 
Quantum fluctuations are redshifted to exit the cosmological horizon one after another. 
From the static observer's point of view, nothing goes out of the horizon 
but the conformal zero mode accumulates at the horizon. 
We build on the stochastic picture of infrared (IR) fluctuations \cite{Starobinsky1986,Tsamis2005}. 
It is closely related to the diffusion process of the conformal zero mode.
The resummation of IR logarithmic effects in quantum gravity, 
which are $O(\log ^na)$ in terms of the scale factor $a$ of the universe 
play an essential role in cosmic expansion\cite{Tsamis1994,Kitamoto2012}. 

In our study of IR effects in quantum gravity, Einstein gravity plays the central role as the mass-less sector of string theory contains Einstein gravity. Some modifications of Einstein gravity are proposed in connection with
inflation and dark energy. Phenomenologically it is necessary to introduce a single scalar field (inflaton)
which couples minimally to gravity with a potential. By satisfying a slow roll condition, it can expalin
the essence of inflation . Although it is a classical modification of the Einstein gravity,
the quantum effect is at the heart of the theory, i.e. the vacuum fluctuation is stretched beyond Hubble scale by  inflation.
It should be possible to construct fully quantum theory to supersede these semi-classical theories.

There is another possibility that classical and quantum descriptions are dual to each other. 
%In Appendix A, we have summarized  evidences for de Sitter duality by a perturbative analysis. We have adopted a BRS type local gauge.
We have computed the one loop quantum correction to Einstein gravity. It is dual to an inflation theory
in the sense that we obtain the former by integrating out the inflaton from the latter. 
We have further constructed a pair of such theories with  scale invariance   in Appendix B.  
They are candidates of quantum gravity at the UV fixed point.
%The quantum gravity at the UV fixed point is obtained by integrating out the inflaton in a classical inflation theory. 
%With large anomalous dimensions, this is an example of fully non-perturbative duality.:
The gravity has always been investigated from geometric point of view. Inflation theory is no exception.
On the other hand, that approach leaves too many freedom. What is inflaton ? Why they perform slow roll
? and so forth.
There is an alternative  stochastic point of view. This approach relays on more microscopic point of view.
Inflaton coud be understood as quantum effects. More systematically, stochastic equations  on the boundary play the major role like the Einstein's field equation.
We derive Fokker Planck equation as the major apparatus to explore the universe. 
This dual approach has instant rewards. Namely, the mysterious slow roll is just the random walk
of conformal mode. 
Furthermore the slow roll parameter $\epsilon$ is predicted to be $O(1/N)$ which is the universal fractal dimension ,of the random walk. In fact FP equation becomes as important as the Einstein's equation.
We derive FP equation from the renormalization group.
It can determine the evolution of the entropy of the universe.
So each solution is a history of the universe.  

In de Sitter space, the scale invariance of the infra-red quantum fluctuations may give rise to time dependent effects such as shielding the cosmological constant. 
We suspect that  the slow roll of inflaton
might arise  in quantum gravity without introducing the potential by hand.
As $(Ht)^n\sim (\log a)^n$ grow during the long cosmic history,
we need to resum IR logarithms to all orders in order to demonstrate such a mechanism.
The renormalization group is one of the most powerful tool to accomplish such a feat .
In the critical phenomena, the divergence of the correlation length is caused by IR divergences.
The imporant point is that the only the massless scalar modes  play the significant role as explained in Appendix A. We call it  $X$ field. It is nothing but the curvature perturbation $\zeta$ and has the finite overlap with others $\omega=(\sqrt{3}/4)X,~~h^{00}=h^i_i =2\omega $ .

The stochastic equations can be solved exactly under Gaussian approximation. We have found the 2 types of solutions; (a):UV complete spacetime and (b): inflationary spacetime with the power potentials.
Inflation theory is a classical gravitational theory. It provide a geometrical description of de Sitter like space-time. Although we derive GFP from microscopic method, the identical GFP equation follows from the classical thermodynamics in the inflation theory. It is because the von-Neumann (entangled) entropy of the two sub-sectors of the Hilbert space agree.
$S_A=S_B$.
%This fact offers a strong evidence for de Sitter duality.
Its solution with the ultraviolet fixed point indicates that 
the Universe starts with the de Sitter expansion near the Planck scale with $\epsilon=0$. 
We argue the inflationary universe subsequently dominates to maximize the entropy. The pre-inflation era
may be necessary to prepare launching inflation era.

In order to elucidate the IR logarithmic effects nonperturbatively, 
we formulate a Fokker Planck equation for the conformal zero  modes of the metric. 
We obtain the $\beta$ function for dimension-less gravitational coupling $g=G_NH^2/\pi$ in a Gaussian approximation . $G_N$ is the Newton's coupling and $H$ is the Hubble parameter. It is valid for a small coupling $g$ in quantum gravity except at the beginning of the Universe. 
We first sum up the $O(\log ^na)$ terms to all orders to identify the one-loop running coupling $g=2/\log N$.
We next resum the $1/(\log ^nN)$ type corrections to  take account of  the quantum back-reaction on $g$. 
Since the $\beta$ function turns out to be negative, 
it implies asymptotic freedom for $g$ toward the future \cite{Gross1973,Politzer1973}. 
Furthermore, the $\beta$ function possesses the ultraviolet (UV) fixed point in the past with the critical coupling $g=1/2$. 
This fact indicates that our Universe begun with the dS expansion near the Planck scale with a minimal entropy.

Since $\epsilon\propto \beta$, the  smallness of
$\epsilon$ in the beginning of the Universe is naturally explained. 
Strictly speaking, our approximation is justifiable when the coupling is small. 
Nevertheless, $g=1/2$ might be small enough such that our picture qualitatively holds.
In our approach,
the quantum correction to de Sitter entropy is given by the von Neumann entropy of the zero mode. We may decompose the total Hilbert space  into the bulk  $H_A$ and boundary $H_B$.
$H=H_A\otimes H_B$. 
The de Sitter entropy is von Neumann entropy,i.e.  entangled entropy. 
Quantum definition involves the integration on A: $ \rho_B=Tr_A\rho$
This operation corresponds to the renormalization of $\rho$ at the horizon scale.
Then on B:
$S_B=-Tr_B\rho_B\log(\rho_B)$
We have counted the entropy of the lightest (mass less) sector
 around the cosmological horizon. 
 It is equal to bulk entropy: $S_A$.
 In this sense it may not be surprising that holographic $\beta$ function 
 predicts UV fixed point.

What is de Sitter duality?  That is the question we have addressed.
de Sitter like space-time have dual facets. 
Inflation theory with slow roll inflaton  presents
classical geometric view. In contrast stochastic point of view
 is equally convincing as it explains the slow roll parameters
 in terms of fractal dimensions of random walk. 
 $1/\epsilon\sim N$ holds for every inflation theory with power potentials.
 In the stochastic picture, it is simply understood conformal mode  is performing the random walk. So inflationary universe represents a particular universal class.
 
 It takes time to get the destination
 by random walk.  Thus stochastic equations are as important as Einstein's
  equation.
 The major achievement of this paper is to derive GFP equations from the IR quantum effects in Einstein gravity.  This equation can be justified also from geometric/thermodynamic point of view.
 So we have established geometry in the bulk/stochasticity on the boundary.
 We have precursors to our present understanding of de Sitter duality such as perturbative equivalence between inflation theory and Einstein gravity.
They may even make sense at the UV fixed point.   They are reported in Appendix A and C.

The GFP relates the evolution speed 
of $S$ and the scale of the universe. We can reconstruct the history of $S(t)$ by integrating GFP. In other words, a solution of GFP is a history of the universe.
There are the UV finite solution and inflationary solutions in our fundamental equation.

In section 3, we argue the random walk of conformal zero mode is dual to slow roll inflation: This intuition has been our main impetus to pursue dS duality .
In section 4, 
we derive the GFP from classical thermodynamics for inflation theory in a dual picture  to the original quantum stochastic picture. It is base on the relation
$S_A=S_B$.
It justifies  de Sitter duality at the non-perturbative level.
In section 5, we conclude this paper  by listing our successes so far and tasks remaining ahead.

We have worked for sometime on the de Sitter space. We have adopted the BRS type local gauge.
In appendix A, we review our set up of the local perturbation theory .  We may recall our convention
anytime there.
We also summarize the IR behavior of the Einstein gravity by an effective theory. 
 We renormalize distribution function in Appendix B.
 In Appendix C, we report our candidate for the holly grail of quantum gravity:
the scale invariant quantum gravity and its dual inflation theory.
Appendix D is a brief summary of stochastic equations to make this paper self-contained.
Appendix E is devoted to identify the strong/weak coupling regime
in comparison to observational prospects.
%We investigate composite (pre-)inflationary universes in Appendix E.

The major results of this paper are the derivation of the GFP by 
renormalizing the distribution of conformal mode and the
investigations of its solutions: the UV finite and inflationary space-time. 
We further derive it in a dual geometric picture, extending de Sitter duality between Quantum( bulk)/Stochastic (boundary) gravity at non-perturbative level.
We argue there are pre-inflationary era prior to the inflationary era.
Firstly the UV finite solution indicates the universe was born as an almost Planck scale de Sitter
space. The gravitational- coupling decreaces logarithmically with time.
$\epsilon$ stays small before the inflation takes place. It ensures that the universe undergoes sufficient inflation to initiate big bang.
In other words, we propose there is a pre-inflation era which is necessary to solves $\epsilon$ problem.

\section{Renormalization group and stochastic universe}
%\setcounter{equation}{0}
%\subsection{Equation \numbers}

Gibbons and Hawking pointed out that de Sitter space has a geometric entropy 
proportional to the area of the horizon \cite{Gibbons1977}. 
They rotate $dS_4$ into $S_4$. The volume of $S_4$ is $V_{S^4}={8\pi^2 / 3H^4}$. The Einstein action on $S^4$ is
$V_{S^4}{6H^2 / 16\pi G_N}=1/g$. de Sitter entropy is the inverse of the dimensionless coupling $g=G_NH^2/\pi$.
We may consider de Sitter like space-time which undergoes the slow roll inflation.
If the Hubble parameter decreases with time, 
the dS entropy must increase simultaneously.
We investigate four-dimensional gravity on dS space from an entropic point of view. 
We have postulated that the quantum correction to the dS entropy comes from the scale distribution of the universe
\cite{Kitamoto2019-1,Kitamoto2019-2}. 

This conformal mode distribution operator is renormalized due to IR logarithms. For example it may acquire time dependence.
We prove that it obeys Fokker-Planck equation.
We subsequently solve the FP equation and identify different types of the  solutions.
This is the major goal of this paper.
By applying the renormalization group, we derive the Fokker-Planck and Langevin  equations  for the conformal mode distribution function  $\rho(\omega)$ at the horizon scale.  
We emphasize that the boundary theory  is very different from the bulk theory. 
The bulk theory is quantum field theory or its UV completion like string theory.
The FP equation implies the boundary theory is stochastic. 
It establishes the stochastic nature of conformal zero mode $\omega$ and
the curvature perturbation $\zeta$ at the horizon exit.
In contrast, there is no universality concerning  shape of the potential .
	
Suppose $\zeta$ is made of the pile of the modes $ \zeta =\sum_i \Delta\zeta_i$
where $i $ denotes different horizon exit time. 
The constituants have non-vanishing self-correlation . $<\zeta^2>=\sum_i <(\Delta \zeta_i)^2> 
=4gN/m=g/\epsilon$ for curvature perturbation. The magnitude of the two point function is proportional to the e-folding number $N$. This represents the  identical fractal dimension 2 with the random walk.
The slow roll parameter is $\epsilon\sim O(1/N)$ .	  The power of the inflaton potential $m$ does not change this ubiquitous signal although it changes the magnitude of the 2 point function.

We investigate the distribution of the conformal zeromode $\omega$ . It is assumed to be Gaussian since gravity is weakly coupled except near the Planck scale. We start with the Gibbons-Hawking type distribution. We focus on the massless Lorentz scalar mode. There is only one such a mode $\omega$ in Einstein gravity in our gauge. In fact $\omega\sim \sqrt{3}/4 X$ after projecting out massive mode.
$X$  is equal to   curvature perturbation $\zeta$. $X$ has negative norm and the rest of the variables with non vanishing mixing will inherit it. 
 This part is explained in Appendix A for self-contained-ness  and establish the standard convention.

Let us consider 2 point function of the conformal mode at the coincident limit. The conformal modes are creation and annihilation eigenstates but not the mass eigenstates . We project the conformal mode onto $X$ field and let $X$ field propagates just like quarks or neutrinos. So the propagator of $\omega $ has the projection factor $T=<\omega|X>,|T|^2=3/16$.

dS space may be decomposed into the bulk and the boundary, i.e., the sub-horizon and horizon. 
From a holographic perspective, we consider the conformal zero mode dependence of the Einstein-Hilbert action: 
\begin{align}
\frac{1}{16\pi G_N }\int \sqrt{g}d^4x(Re^{2\omega}-6H^2e^{4\omega}) 
\sim\frac{\pi}{G_N H^2}(1 -4\omega^2), 
\label{action1}\end{align} 
where the gauge fixing sector is suppressed due to no renormalization theorem.
Our linear gauge fixing procedure and the propagators are explained in Appendix A.

The quadratic part of $\omega$ constitutes a Gaussian distribution function for the conformal zero mode, 
\begin{align}
\rho(\omega)=\sqrt{\frac{4}{\pi g}}\exp\big(-\frac{4}{g}\omega^2\big).
\label{distribution1}\end{align}
It may represent an initial state of the Universe when the dS expansion begins. 
In order to describe the time evolution of the conformal factor of the Universe, 
we make the gravitational coupling time dependent $g(t)$
and introduce a new parameter $\xi(t)$.
\begin{align}
\rho(\xi (t),\omega)=\sqrt{\frac{4\xi (t)}{\pi g(t)}}\exp\big(-\frac{4\xi (t)}{g(t)}\omega^2\big). 
\label{distribution2}\end{align}

$\xi$ is the only parameter in the Gaussian approximation. Note $g(t)$ is fixed by the two point function of the  tensor mode. We find $g(t)$ must be time dependent although it changes much slower $O(1/\log N)$ than $\xi\sim O(1/N)$ in the large $N$ expansion.
%In what follows, $\xi, g$ are assumed to be time dependent unless announced otherwise.
We work within the Gaussian approximation since it is an excellent approximation 
for gravity with the small coupling $g$.
The entropy of the distribution $S=-\text{tr}(\rho \log \rho)\sim {1/2\log (g/\xi )}$ becomes larger as $\xi$ becomes smaller. Thus the diffusion triggers an instability in de Sitter space.

 This is the set up of this work. $\xi$ is introduced to describe inflationary universe. 	If the universe evolves slowly, it must be described this way with a slowly changing deformation parameter $\xi$.
We can measure possible fractal dimension
of the universe from the distribution function of the conformal mode. 
As we prove that it obeys FP equation, $\xi$ is a necessary parameter for the solution.
It implies the conformal zeromode performs random walk. We claim the slow roll relation $\xi\sim O(1/N)$ is a clear signal for this fact . This is very universal phenomena observed in every inflation model with the power potential. Our motivation to study de Sitter duality stem from the recognition of this amazing simplicity behind the mysterious slow roll.
The major purpose of this paper is to show the duality between slow roll in inflation theory  and stochastic physics on the boundary.
This equation controls the evolution of the entropy of the universe $S\sim (1/2)\log ( g/\xi)$ which is carried by conformal zeromode.
It has rich spectrum of the solutions and they may have strange and beautiful stories to tell.

The condensation of Lorentz singlet field has been scientifically accepted .
Lorentz symmetry is gauged in quantum gravity.
Since gravity is weak coupling unless around the Planck scale, 
Lorentz symmetry should appear as global symmetry in agreement with experiments.
In fact particles form representations  of the Poincare group.
The vacuum must be the singlet.
In terms of  the distribution function, the $n$-point functions are defined as follows 
\begin{align}
\langle\omega^n(t)\rangle=\int d\omega \rho(\xi(t),\omega)\omega^n. 
\label{n-pt}\end{align}
As explained in Appendix C, (\ref{distribution2}) is a solution of the FP equation with no drift force, equivalently a solution of  diffusion equation with 
$\xi(t)=1/6N(t)$ where $N(t)=\int^t dt' H(t')$ is the e-folding number.
$1/\xi$ is the enhancement factor of the conformal perturbation. 
In slow roll inflation theory, $\xi=3\epsilon/2$. This equation holds also for the linear potential.
We determine the relation for more generic case in section 4. 
Such an enhancement arises by the diffusion of the distribution $\rho $.
Since the mathematics is the same with the random walk, the physics
must be the same. The universe becomes larger by the jolts of horizon crossing
conformal modes.  

$1/\xi$ is proportional to the e-folding number $N$. This is the universal feature among slow roll single field inflation theory.
It is certainly important to recognize the hidden identity.
Our fundamental realization to advocate  de Sitter duality is that the factor $O(N)$ enhancement of the scalar  to tenor ratio is
the universal prediction of FP equation.
The negative norm of the bulk conformal mode is crucial to
obtain the consistent FP theory as explained in the next section.
It is the reason that the cosmological constant decreases toward the future.

 We derive FP equations from many view points. We first adopt the microscopic approach. We derive FP equation from renormalization group. We then derive 'it from macroscopic thermodynamics. We finally check the consistency with Langevin formalism.

We investigate the dynamics of conformal zero mode after integrating the bulk modes.
The two point function gives rise to IR logarithms.
\begin{align}
<\omega^2>_{\text{bulk}}&=-\int_{k^*}\frac{dk}{k} \frac{3g}{4}
\notag\\
=& \frac{3g*}{4}(\log(k^*))
=\frac{3g^*}{4}N*.
\label{IRlog}\end{align}

We recall the following identity holds at the horizon exit $t=t^*$
\cite{Maldacena203}.
\begin{align}
\dot{N}^*\exp({N^*})=k.
\end{align}
where $*$ denotes the exit time.
 It is the reason why the renormalization scales are related $\log k = Ht^*$ in
 (\ref{IRlog}). It is the first come, the first served system. 
 The softer the plain wave, the earlier its exit time.

Note the norm is negative,but $k$ is the exiting momenta at the horizon and it increases with exiting time $t^*$. So the above quantity increases with time due to double negation. Fortunately, negative norm of the conformal mode leads to the correct FP equation.

We need to use time independent UV cut-off like dimensional regularization to justify this argument.
This is the second term of the Schwinger-de Witt expansion of the propagator.
It is logarithmically divergent.
\begin{align}
 -\frac{3}{16\pi}G_N({R\over 6})(\frac{2}{\epsilon}-{2\log(k_*)}).
 \label{SDT}
\end{align}

The leading term is quadratically divergent but should be subtracted
as it vanishes in dimensional regularization.
We assume UV cut-off has no time dependence. In $D=4-\epsilon$ dimension,
it has $1/\epsilon $ pole . After the subtraction of this pole, we find (\ref{IRlog}). 
%For simplicity, we have assumed  that $g,H$ are constant during inflation.
%In practice, we can consider local effects to avoid artificial errors.

 %$t_e$ denotes the time when the inflation is completed.
As we explained, there is no time dependent UV contributions. 
We focus on the Hubble scale physics where $a(t)=1/(-\tau H)=\exp(Ht)$
and $|k\tau|\sim  1$ ,
Our renormalization scale is
$\log k\sim-\log (-\tau )\sim Ht=N$.
While the wave functions of the bulk modes oscillate with respect to $\tau$,
the boundary mode is constant. That is why we call it the zero mode.
Our strategy is to integrate out oscillating modes first.

We thus construct low energy effective 
theory around the Hubble scale by renormalizing the distribution function.
This is the high-light of this paper. The technical details are delegated in Appendix B to streamline the main story.
The bare distribution function is given by subtracting the bulk mode contributions . It is independent from 
the renormalization scale or exit time because they are subtracted simultaneously with the UV divergences.
The subtraction of UV divergences of $\rho$ is accomplished by multiplying the following operator
from the left

\begin{align}
\rho_B=\exp\big(\tilde{g}(t)(\frac{1}{2\epsilon}-\frac{Ht}{2})\frac{\partial^2}{\partial\omega^2}\big)\times\rho .
\label{bare}\end{align}
As $\rho_B$ is independent of the exiting time, $\dot{\rho_B}=0$,
the distribution function $\rho$ obeys the following renormalization group equation,: 
\begin{align}
{\partial\rho\over \partial N}
=\frac{3g(t)}{4}\frac{1}{2}\frac{\partial^2\rho}{\partial \omega^2}
\label{FP0}\end{align}
The factor $3/16$ in the diffusion term is the projection factor  $|<X|\omega >|^2$. 

We neglect conformally coupled modes because it has the effective mass $m^2=2H^2$. 
It cancels 
 the coefficient of  the logarithmic contribution in (\ref{SDT}) as $R/6=2H^2$. There is no drift term in the reduced space consisting only of $X$. 
In such a subspace, $\omega =\sqrt{3}X/4 $ approximation is well defined.
The gravitational FP equation: GFP (\ref{FP0}) is  obtained by
integrating  the  quantum bulk modes inside the horizon. It turns out to be a diffusion equation due to the lack of the drift term.

We emphasize the bulk and boundary theories are completely different.
Although the Bulk theory is the quantum gravity,  the boudary
theory is the stochastic FP equation. The analogy is that the negative  sign of $A_0$ propagator gives the correct Coulomb force in QED.
The solution of FP equation implies the universe expands like the random walk  jolted by the horizon exiting conformal modes. The universal prediction is the slow roll parameter $\epsilon \sim 1/N$.
We can sum up the IR logarithms $\log ^na=(Ht)^n$ by this equation to find a running coupling $g(t)$. 
We first work out the leading log solution.
%We subsequently obtain the exact solution.

The leading log approximation is valid in the large N limit.
The FP (diffusion) equation  shows that the solution is the Gaussian distribution with the  standard deviation  $O(1/N(t))$ as shown in the Appendix C \cite{Parisi}. 
We first check whether we are on the right track.
 The distribution entropy of the conformal mode  increases logarithmically, 
\begin{align}
S=\frac{1}{2}\log {1\over\xi}
\sim\frac{1}{2}\log N(t). 
\label{entropy1}\end{align}
%where we have kept the leading $log$ part.
In de
Sitter space, entropy may be identified with the effective action 
with the opposite sign. We maximize the entropy instead of minimizing the effective action.

Identifying the entropy of conformal zero mode
with the quantum correction to dS entropy,
we obtain the bare action with the counter term
\begin{align}
\frac{1}{g_B}=\frac{1}{g(N)} - \frac{1}{2}\log(N). 
\label{coupling}\end{align}
By requiring the bare action is  
independent of the renormalization scale: namely  $N$ , as we have just done to derive FP equation,'
we obtain the one loop $\beta$ function.
\begin{align}
\beta=\frac{\partial}{\partial \log(N)}g(N)=-\frac{1}{2}g(N)^2. 
\label{beta1}\end{align}
%It is valid at the one loop level although higher loop effects have been ignored.
We find the running gravitational coupling in the leading log approximation:
%\footnote{The coefficient in front of $N$ can be put to the identity.}
\begin{align}
 g(N)={2\over \log(N)}. 
\label{beta1}\end{align}
Semiclassical
de Sitter entropy is in agreement with this estimate
(\ref{entropy1}).
\begin{align}
S(N)={1\over g(N)}={\log(N)\over 2}. 
\label{beta1}\end{align}

We have collected convincing support on our postulate. The semiclassical evidence is that the both are  given by the gravitational coupling $1/g=\log N/2$ )We show the renormalized distribution function obeys gravitational Fokker-Planck equation (GFP) and Langevin equations .
Under the Gaussian approximation, they boil down to a simple first order partial differential equation. 
The identical equation is derived by the thermodynamic arguments in the inflationary space-time. GFP determines the evolution of de Sitter entropy of the universe. It coincides with $\beta$ function of $g$.

The holographic investigation at the boundary shows that 
$g$ is asymptotically free toward the future. 
The renormalization group trajectory must reach Einstein gravity in the weak coupling limit 
for the consistency with general covariance \cite{Kawai1993}. 
We find that it approaches a flat spacetime in agreement with this requirement.

We have evaluated the time evolution of entropy to the leading log order in (\ref{FP0}). 
In order to take account of the higher loop corrections in $g$, the FP equation 
should be precisely formulated. It turns out to be just necessary to make the equation covariant and local:
\begin{align}
{\partial\over \partial N}{\rho_N}
-\frac{3g(N)}{4}\cdot\frac{1}{2}\frac{\partial^2}{\partial \omega^2}\rho_N
=0.
\label{GFP}
\end{align}
The covariance and locality ensures that
$1/\epsilon$ pole induced $t$ dependence  cancel just analogously.

\section{Perturbative and UV complete solution of GFP}

The solution of FP equation is given by the Langevin formalism in Appendix C
\begin{align}
<(\omega(t)-\omega(0))^2>
=\int_0^t dt' \frac{3g(t')}{4}H(t')
=\int_0^N dN' \frac{3g(N')}{4}.
\end{align}
\begin{align}
{\partial\over \partial N}<(\omega(N)-\omega(0))^2>
%=\int_0^t dt' \frac{3g(t')}{4}\cdot\frac{1}{2}H(t')
= \frac{3g(N)}{4}.
\end{align}

The identical two point function is obtained from FP equation (\ref{GFP}).
\begin{align}
{\partial\over \partial N}\int d\omega\rho_N(\omega)\omega^2
=&\frac{3g(t)}{4} \int d\omega\frac{1}{2}\frac{\partial^2}{\partial \omega^2}\rho_N(\omega)\omega^2\notag\\
=&\frac{3g(t)}{4}\int d\omega \rho_N(\omega)%=&\int d\omega \frac{3g(N)}{4}\cdot\frac{1}{2}\rho_N(\omega)\notag\\
= \frac{3g(t)}{4}.
\label{FP2p}\end{align}

In what follows, we consider the following Gaussian solutions of FP equation.
\begin{align}
\rho=\sqrt{\frac{4\xi (t)}{\pi g(t)}}\exp\big(-\frac{4\xi (t)}{g(t)}\omega^2\big).
\label{gausaz}
\end{align}
This distribution function determines the two point function of the conformal zero mode.
\begin{align}
\int d\omega \omega^2 \rho(\omega) ={g(t)\over 8\xi(t)}.
\end{align}
 We put the Gaussian ansatz into the FP equation (\ref{FP2p}) and
find the following condition for the background to satisfy.
\begin{align}
{\partial \over \partial N}
\left({g\over 8\xi}\right)={3g\over 4}.
\end{align} 

We thus obtain a  simple equation in terms of e-foldings.
\begin{align}
\frac{\partial}{\partial N}\log \frac{g(t)}{\xi}=6\xi. 
\label{FP3}\end{align}
This is the major result of this paper. Since the  entropy
of the conformal mode is $(1/2)\log \frac{g(t)}{\xi} $,
this equation determines the evolution of entropy of a universe.

As we have emphasized, Gaussian approximation must be valid unless 
we probe Planck scale.
(\ref{FP3}) determines the evolution of von Neumann entropy $S={1\over 2}\log{g\over \xi}$ with respect to $N$.
This formula  confirms the validity of our postulate that distribution entropy
of conformal zero mode constitutes the quantum correction to de Sitter entropy.

Secondly, the leading log approximation is valid in the large $N$ limit
since $g\sim O(\log N)$ can be regarded as a constant.
We obtain the equation which is the large $N$
limit of (\ref{FP3}).
Equivalently we put it as follows.
\begin{align}
{\partial\over \partial N}{\log \xi}=-6\xi.
\label{FP4}
\end{align}
The solution is
\begin{align}
\xi=\frac{1}{6N}. 
\label{xisol}\end{align}
Here we have neglected time dependence of $1/g \sim \log N$
in comparison to $1/\xi\sim N$.
In such a limit, (\ref{FP3}) turns into
(\ref{FP4}). 
For finite $N$, $\xi=1/6N$ will acquire finite $N$ corrections like
 (\ref{solution1}).

As it happens $\epsilon=1/4N$ for the linear potential, the result is in agreement with the slow roll picture $<\zeta^2> =g/\epsilon$. In this estimate, we have used Gaussian ansatz and large $N$
expansion. It appears that non-perturbative approach is necessary to probe generic potentials.
The solutions with more generic potentials are considered in the next section from  de Sitter duality point of view.
Our basic conjecture is that the mysterious slow roll of inflaton represents the Brownian motion of the conformal zero mode. 
This idea explains generic features and may develop  further to demystify  de Sitter duality.
The attractive point of our theory is its simplicity. 
The inflaton is replaced by the random walk of the conformal zeromode. The universal enhancement of scalar to tensor ratio by $O(N)$ is explained by its fractal dimension. 
However the precise coefficient depends on the details.

In the literature, $\delta N$ formalism is widely used to study the curvature perturbation. It underscores the validity of the stochastic picture of the inflation
\cite{Starobinsky1985},\cite{Bond,Stewart,Lyth2005a,Lyth2005b}.
 Let us consider the fluctuation of the curvature perturbation $\zeta$.
\begin{align}
{ \zeta} =&{\delta N} ={H\over \dot{ f}}\delta f ,\notag\\
<\delta\varphi(t) \delta\varphi(t')>& =({H^2\over 4\pi^2})H\delta(t-t').
\label{Lgdn}
\end{align}
$\delta \varphi$ obeys the Langevin equation.

We obtain in the super-horizon regime:                                                                                                                 
\begin{align}
&<\zeta^2(t)> =<  ({H\over \dot{f}})^2(\delta f)^2>\notag\\
&={1\over 2\epsilon M^2_P}<(\delta f )^2> \notag\\
&=
\int ^{N}dN'{H^2
\over 8\pi^2\epsilon M^2_P}.
\label{P123}
\end{align}
Thus
\begin{align}
{\partial \over \partial {N}} <\zeta^2(t)> =	{g\over \epsilon}=P.
\label{zeta2}
\end{align}
There still remains unexplained parameter $\epsilon$ in this approach. 
In contrast, $\epsilon$ is  predicted to be $O(1/N)$ in our stochastic equation.

There is a UV fixed point in our renormalization group. %We briefly study it.
FP equation (\ref{FP3}) enables us to evaluate higher order corrections to the $\beta$ function. 
The expansion parameter is $1/\log N$. 
We can confirm that the following $g_f$ and $\xi_f$ satisfies (\ref{FP3}), 
\begin{align}
g_f=\frac{2}{\log N}\big(1-\frac{1}{\log N}\big),\hspace{1em}
\xi_f=\frac{1}{6N}\big(1-\frac{1}{\log N}\big). 
\label{solution1}\end{align}
The leading log results are (\ref{beta1}) and (\ref{xisol})  respectively.
Thus, the $\beta$ function, $\epsilon$ and the semi-classical entropy generation rate are given by 
\begin{align}
\beta=\frac{\partial}{\partial \log N}g_f=
 -{2\over \log^2 N}+{4\over \log(N)^3}.
\label{beta2}\end{align}

\begin{align}
\epsilon_f = -{1\over 2}{\partial \over \partial N}\log(g_f)=
 -{1\over 2g_fN}\beta_f.
\end{align}
\begin{align}
\frac{\partial}{\partial N}S_{sc}=\frac{\partial}{\partial N}\frac{1}{g_f}
=-\frac{1}{Ng_f^2}\beta_f
\label{entropy2}\end{align}

A remarkable feature is that the coupling has the maximum value $g=1/2$ at the beginning. 
It steadily decreases toward the future 
as the $\beta$ function is negative in the whole region of time flow. 
It has two fixed points at the beginning and at the future of the Universe. 
The existence of the UV fixed point may indicate the consistency of quantum gravity. 
The single stone solves the
$\epsilon$ problem \cite{Penrose1988} as well since it vanishes at the fixed point. 
The $\beta$ function describes a scenario that 
our Universe started the dS expansion with a minimal entropy $S=2$ while it has $S=10^{120}$ now. It corresponds to $N=e^2\sim 7.4,  a=e^N\sim 1.6 \times 10^3$. 
The just born Universe is rather large which reflects the critical  coupling $g=1/2$ is rather small. In terms of the reduced Planck mass, $H^2/{4\pi}^2 M_P^2=1$. 

Since we work with the Gaussian approximation, our results on the UV  fixed points are not water tight
as the coupling is not weak. Nevertheless we find it remarkable that they
support the idea that quantum gravity has a UV fixed point 
with a finite coupling. In fact 4 dimensional de Sitter space is constructed in the target space 
at the UV fixed point of $2+\epsilon$ dimensional quantum gravity  \cite{Kawai1993}.
4 dimensional de Sitter space also appears at the UV fixed point of the exact
renormalization group \cite{Reuter}\cite{Souma}.

Such a theory might be a strongly interacting conformal field theory .
However, it is not an ordinary field theory as the Hubble scale is Planck scale. 
We have cited a pair of the candidate theories as an example of de Sitter duality in Appendix C.
Our dynamical $\beta$ function is closely related to the cosmological horizon and physics around it. 
The existence of the UV fixed point could solve the trans-Planckian physics problem. 
A consistent quantum gravity theory can be constructed 
under the assumption that there are no degrees of freedom at trans-Planckian physics \cite{Bedroya2019}. 
In this sense, it is consistent with string theory and matrix models.  
The Universe might be governed by (\ref{solution1}) in the beginning 
as it might be indispensable to construct the UV finite solutions of the FP equation. 

\section{Inflationary universe emerges with power potentials}

As it turns out, the equation (\ref{FP3}) has another class of solutions: the inflationary universe  with the power potentials.
\begin{align}
g=c\tilde{N}^\frac{m}{2},\hspace{1em}\xi=\frac{m+2}{12\tilde{N}}.
\label{solution2}\end{align}

Here $c$ is an integration constant.
$m$ denotes the power of the potential: $f^m$.
We have changed the variables by
replacing $N$ by $\tilde{N}$
where $\tilde{N}=N_e-N$.  $N_e$ denotes the e-folding number at the end of inflation.  To be precise, they are the solutions of the following equation:
%the time reversal of (\ref{FP3})
\begin{align}
-\frac{\partial}{\partial N}\log \frac{g(\tilde{N})}{\xi(\tilde{N})}=6\xi(\tilde{N}). 
\label{FPtr}\end{align}

 (\ref{solution2}) shows  $\epsilon$ and $\xi$ are related as 
\begin{align}
\epsilon=\frac{m}{4\tilde{N}}=\frac{3m}{m+2}\xi.
\label{epxi}
\end{align}

The inflaton may be identified with the stochastic variable $f$ whose correlators show characteristic features of Brownian motion.  $<f^2>=\tilde{N}$ and $g\propto \tilde{N}^{m/2}=<f^m>$. 
The increase of the entropy $S=1/g$ can be evaluated by the first law $T\Delta S=\Delta E$ 
where $\Delta E$ is the incoming energy flux of the inflaton  \cite{Frolov2003}.
Obviously we are estimating the entropy of super-horizon.

In this way, the one of the Einstein's equation is obtained:
\begin{align}
\dot{H}(t)=-4\pi G_N \dot{f}^2.
\end{align}
This formula can be re-expressed 
in terms of the slow roll parameter: $\epsilon=\dot{f}^2/2H^2$
\begin{align}
2\epsilon=-{\partial \over \partial {N}}\log g(t).
\end{align}
We add $O(1/\tilde{N})$ quantity to the both sides of the equation.
\begin{align}
2\epsilon+{1\over \tilde{N}}=-{\partial \over \partial {N}}\log (g \tilde{N}).
\label{mf2}
\end{align}
We find the relation between  $\xi$ and $\epsilon$
in (\ref{epxi}) : $6\xi=2\epsilon+(1/\tilde{N})=(m+2)/2\tilde{N}$ is reproduced.
We have thus shown that  curvature and conformal perturbation are identical up to the constant factor
(\ref{solution2}).

For power potential inflationary universe, (\ref{mf2}) is rewritten as
\begin{align}
6\xi={\partial \over \partial {N}}\log {g\over \xi}={\partial \over \partial {N}}\log {g\over \epsilon}.
\end{align}
It is consistent with our GFP (\ref{FP3}). The second equality arises since $\xi/\epsilon$ is independent of ${N}$ as shown in (\ref{epxi}). This ambiguity
corresponds to a constant $c$ in (\ref{solution2}). 
In other words,  $\xi={m+2}/{12\tilde{N}}, \epsilon={m/ 4\tilde{N}}$ are $O(1/\tilde{N})$. The both can balance the equation.

In this form, it is evident that $6\xi$ is the scalar spectral index in agreement with slow roll inflation theory. Tensor to scalar ratio is $16\epsilon$ .
We have shown GFP can be derived by macroscopic arguments also.
From quantum gravity point of view, this is an important non-perturbative evidence for 
de Sitter duality.
\begin{align}
\epsilon &= -{1\over 2}{\partial \over \partial N}log(g)=\frac{m}{4\tilde{N}},\notag\\ 
&1-n_s=6\xi=\frac{m+2}{2\tilde{N}}.
\label{infsol1}\end{align} 
In this case, the concave power solutions can be obtained by formally replacing $m$ by $1/n$. We admit it is not entirely clear why concave potentials are relevant.  It is possible that
the convex potentials are already excluded by observations.

In contrast, there is more room for concave potentials to
accommodate the observational information with the judicious choice of $n$.
\begin{align}
%\begin{center}
\epsilon &=-{1\over 2}{\partial \over \partial N}\log(g)=\frac{1}{4n\tilde{N}} ,\notag\\
&1-n_s=\frac{\frac{1}{n}+2}{2\tilde{N}}.
%\end{center}
\label{infsol2}
\end{align}
In our view the inflaton is performing the Brownian motion.
Its trajectory is made of infra-red fluctuations.
Nevertheless the convex and the concave potential seems to devide
the inflation models into the two types.
The concave potentials are promising avenue to explore right now.

The comparison (\ref{solution1}) and (\ref{infsol2}) indicates the 
effective couplings are $1/\log {N}$ in the former
\begin{align}
%\xi={1\over 6N}(1-{1\over \log N}), ~~
\epsilon={1\over \log N\cdot 2N} {(1-{2\over \log N})\over (1-{1\over \log N})}.
\label{infsol3}
\end{align}
and $1/2n$ in the latter.)
\begin{align}
 %{{--------*\/\\\\\\\\\\-xi={1\over 6\tilde{N}}(1+{1\over 2n}), ~~ 
\epsilon={1\over 2n \cdot 2\tilde{N}} .
\end{align} 
So concave potentials are in the weak coupling regime while convex
potentials belong to the strong coupling regime.
We believe this corrependence should be taken seriously.

\begin{table}
\begin{center}
\begin{tabular}{ccccc}
&
& $n=1$
& $n=2$
&$n=3$
\\[.5pc] \hline\hline
&$\epsilon $
&${1\over 4\tilde{N}}$
& ${1\over 8\tilde{N}}$
& ${1\over 12\tilde{N}}$
\\[.5pc] \hline
&$1-n_s$
&$\frac{3}{2\tilde{N}}$
&$\frac{5}{4\tilde{N}}$
& $\frac{7}{6\tilde{N}}$
\\[.5pc] \hline
\end{tabular}
\caption{\label{tab:tm1}}
\end{center}
\end{table}

In Table\ \ref{tab:tm1}, we list expected $\epsilon$ and $1-n_s$ in the power potential
model with $f^{1/n},n=1,2,3$.
We note $1-n_s$ is bounded from below by $1/\tilde{N}$ while
$r=16\epsilon $ is not. It is consistent with the current observations.
It is important to establish the bound on $n$.
We recall here the curvature perturbation:
\begin{align}
P\sim {H^2\over (2\pi)^2 2M_P^2\epsilon}\sim 2.2 \times 10^{-9}.
\label{CP}
\end{align}
%$\epsilon=\xi$ when $m=1$ and $\epsilon=2\xi$ when $m=4$. 
It is clearly equal to $<\zeta^2>=g/\epsilon$ generated during inflation.
It is bounded from below $g>10^{-11}$ if $\epsilon>1/200$.

The UV fixed point endows with a special initial condition for the inflation theory.  $\epsilon$ vanishes at the UV fixed point while it grows subsequently (\ref{infsol2}) up to $1/200$ at $N\sim 10$. It then decays again like $1/(2N\log N)$.
Let us investigate pre-inflation scenario.
The entropy grows logarithmicaly $1/g=(\log N)/2$ as $N$ grows after the Universe emerges out of the UV fixed point.
In the inflationary phase, the entropy grows power like $1/g\sim 1/\tilde{N}^{1/2n}$.
The entropy increases slowly  in the pre-inflation era while
it grows much faster in the inflation era. 
Inflation phase inevitably takes over the pre-inflation phase
as the entropy of the former dominates in time.

The simplest scenario assumes that the inflation starts sometime after the 
birth of the Universe . In order to explain the magnitude of the curvature perturbation	, $P\sim 10^{-9}$,
$\tilde{N}\sim 10^{20n}$. Since the inflation era dominates here, a rough estimate  on the timing of the start of the inflation suffices. At that time, $\epsilon$  hit the minimum.
 $1/2N\log(N) \sim 1/(10^{20n}40n)$.
 It becomes as small as $10^{-20}$.
At the beginning of the inflation,
$\epsilon=1/4n\tilde{N} \sim  10^{-20n}/4n \sim 1/(2N\log N)$.
%$\epsilon $ is small.
%since $\epsilon 4n\tilde{N} \sim 1$. 
Finally we reproduce $\epsilon_* = 1/4n\tilde{N}_* $ at the horizon exit.
%So far in this section, we investigated pre-inflation to inflation transition as if it is the second order phase transition. There maybe a different picture. It may be possible to construct 
%UV finite inflationary Universes as composite models. 
%Such models are investigated in Appendix C.

\section{Conclusions and Discussions}
From holographic perspective, 
we have formulated the FP equation for the conformal mode at the boundary, i.e., the horizon. 
The GFP is derived under the Gaussian approximation, 
which may be justifiable for small $g=G_N H^2/\pi $. 
The von Neumann entropy of conformal zero mode gives the quantum correction to the dS entropy $S=1/g$. 
 $g(t)$ undergoes time evolution:
i.e., a quantum-induced inflation picture. 

The FP equation has two types of solutions. 
One of them is the pre-inflation solution (\ref{solution1}) which is UV complete .
%and prepares a special initial condition for inflation by diminishing $\epsilon$ to a tiny value. 
The other is the subsequent inflation solution (\ref{solution2}) which is not UV complete 
but the inflation stops when $\epsilon$ grows up to $1$. 
This transition scenario is supported by the expectation that the solution with dominant entropy is chosen. 
Since the entropy $1/g$ for (\ref{solution1}) increases logarithmically 
and that for (\ref{solution2}) increases in a power-law, 
the former is chosen initially and the latter is chosen if we wait long enough. 

Although it is a persuasive argument, it is another matter to reveal the transition 
process from pre-inflation to inflation theory.   
We have explored  composite solutions which realize the above transition scenario picture.
In the composite solutions, which consist of the logarithmic solution (\ref{solution1}) 
and the the power-law solution (\ref{solution2}) , the entropy $1/g$ always increase with cosmic evolusion.
They may describe the whole history of the Universe from its birth at the UV fixed point.

Although composite Universe has its merits, there remain many questions. First of all, we have constructed approximate solutions only.
We need to clarify wether complete solutions exist or not. Another question is which combination of the solutions account the Universe.
For UV completeness,  the pre-inflation solution (\ref{solution1}) is indispensable. However what principle determines the other component of the Universe, chracterized by the quantum number $n$?
From the entropy considerations, the universe with small $n$ may be favored.
Although there are many interesting issues here, they are beyond the scope of this paper. 

Holography is a specific feature of quantum gravity.
In this paper we have derived the Fokker-Planck equation
on the boundary from the renormalization group equation 
in the bulk. We have proven the de Sitter duality between 
the bulk and the boundary located at the horizon.
de Sitter duality is deeply connected with the von Neumann or entangled entropy.
Suppose the total Hilbert space is factorized into
the bulk states and boundary states. The entropy of the bulk and
the boundary are the same. This explains why de Sitter duality is   important.
It is not expected that $\beta$ function at the UV fixed point
can be calculable from the FP equation at the boundary. 
Nevertheless the dual pair have the identical entropy which
coincide with the effective action.

\section*{acknowledgment}

This work is supported by Grant-in-Aid for Scientific Research (C) No. 16K05336. 
We thank Hiroyuki Kitamoto, Chong-Sun Chu, Masashi Hazumi, Satoshi Iso, Hikaru Kawai, Kozunori Kohri, 
Takahiko Matsubara, Jun Nishimura, Hirotaka Sugawara, and Takao Suyama for discussions.

\appendix
\section{IR renormalization of Einstein gravity}
\setcounter{equation}{0}

For self-containedness, we write down gravitational  propagators in de Sitter space. 
The de Sitter background is given by 
\begin{align}
ds^2=a_c^2(-d\tau^2+dx_i^2),\hspace{1em}a_c=\frac{1}{-H\tau}.  
\end{align} 
We dress the classical solution by quantum fluctuations.

\begin{align}
ds^2=e^{2\omega} \hat{g}(e^h)^{\mu}_{~\nu} dx_{\mu}dx^{\nu}.  
\end{align} 
The quadratic terms in the Einstein-Hilbert action are given by  
\begin{align}
&\frac{1}{\kappa^2}\int d^4x \sqrt{-g}[R-6H^2]\big|_2 \notag\\
=&\frac{1}{\kappa^2}\int d^4x \big[
-\frac{1}{4}a_c^2\partial_\mu h^{\rho\sigma}\partial^\mu h_{\rho\sigma} 
+\frac{1}{2}a_c^2\partial_\rho h^\rho_{\ \mu}\partial_\sigma h^{\sigma\mu} 
+2Ha_c^3h^{0\mu}\partial_\nu h^\nu_{\ \mu}
+3H^2a_c^4h^{0\mu}h^0_{\ \mu} \notag\\
&\hspace{4.5em}-2a_c^2\partial_\mu h^{\mu\nu}\partial_\nu\omega
-8Ha_c^3h^{0\mu}\partial_\mu\omega 
+6a_c^2\partial_\mu\omega\partial^\mu\omega
-24H^2a_c^4\omega^2 \big]. 
\label{EH}\end{align}
We adopt the following gauge fixing term: 
\begin{align}
\int d^4x\mathcal{L}_\text{GF}
&=\frac{1}{\kappa^2}\int d^4x\big[-\frac{1}{2}a_c^2F_\mu F^\mu\big], \notag\\
F_\mu&=\partial_\rho h^\rho_{\ \mu}-2\partial_\mu \omega
+2Ha_c h^0_{\ \mu}+4Ha_c\delta_\mu^{\ 0}\omega. 
\label{GF}\end{align}

The sum of (\ref{EH}) and (\ref{GF}) is given by 
\begin{align}
&\frac{1}{\kappa^2}\int d^4x\sqrt{-g}[R-6H^2]\big|_2+\int d^4x\mathcal{L}_\text{GF} \notag\\
=&\frac{1}{\kappa^2}\int d^4x\big[
a_c^2(-\frac{1}{4}\partial_\mu \tilde{h}^{ij}\partial^\mu \tilde{h}^{ij} 
+\frac{1}{2}\partial_\mu h^{0i}\partial^\mu h^{0i}
-\frac{1}{3}\partial_\mu h^{00}\partial^\mu h^{00} 
+4\partial_\mu\omega\partial^\mu\omega) \notag\\
&\hspace{4.5em}+H^2a_c^4(h^{0i}h^{0i}-h^{00}h^{00}+4h^{00}\omega-4\omega^2) \big], 
\end{align}
where $\tilde{h}^{ij}$ is the spatial traceless mode: 
\begin{align}
\tilde{h}^{ij}\equiv h^{ij}-\frac{1}{3}h^{00}\delta^{ij}. 
\end{align}
The quadratic action is diagonalized as follows 
\begin{align}
\frac{1}{\kappa^2}\int d^4x\big[
&-\frac{1}{4}a_c^2\partial_\mu \tilde{h}^{ij}\partial^\mu \tilde{h}^{ij} 
+\frac{1}{2}a_c^2\partial_\mu X\partial^\mu X \notag\\
&+\frac{1}{2}a_c^2\partial_\mu h^{0i}\partial^\mu h^{0i}+H^2a_c^4h^{0i}h^{0i} 
-\frac{1}{2}a_c^2\partial_\mu Y\partial^\mu Y-H^2a_c^4 Y^2 \big], 
\label{diagonalized}\end{align}
where $X$ and $Y$ are given by 
\begin{align}
X\equiv 2\sqrt{3}\omega-\frac{1}{\sqrt{3}}h^{00},\hspace{1em}Y\equiv h^{00}-2\omega. 
\end{align}

In our parametrization\cite{KKK}, the massless scalar field with the canonical kinetic term is
\begin{align}
X=2\sqrt{3}\omega-{1\over \sqrt{3}}h^{00}.
\end{align}
We may retain  massless minimally coupled $X$ field space:
$h^{00}\sim 2\omega \sim (\sqrt{3}/2)X$.
The curvature perturbation $\zeta$
is the fluctuation of the over all scale of the spacial metric:
 $g_{ij}=e^{2\zeta}\hat{g}_{ij}$.  Indeed we find they are the same 
\begin{align}
2\zeta= 2\sqrt{3}\omega+{1\over \sqrt{3}}h^{00}=2X.
\end{align}

We find no IR logarithms in the BRS trivial sector.
As seen in (\ref{diagonalized}) , 
the Einstein gravity consists of massless minimally coupled modes, 
and conformally coupled modes.   In the latter case, the coefficient
of the second de Witt-Schwinger expansion of the propagator cancels
as $R/6-2H^2=0$.
We neglect the conformally coupled modes 
and focus on the subspace of massless minimally coupled modes 
\begin{align}
h^{00}\simeq 2\omega \simeq \frac{\sqrt{3}}{2}X,\hspace{1em}
\tilde{h}^{ij}.
\end{align}
That is because only the massless minimally coupled modes induce the IR logarithms
\begin{align}
\langle X(x)X(x')\rangle&=-\langle  \varphi(x)
\varphi(x')\rangle \notag\\
\langle \varphi(x)\varphi(x')\rangle&\simeq \frac{\kappa^2H^2}{8\pi^2}\log \big(a_c(\tau)a_c(\tau')\big). 
\end{align}
\begin{align}
\langle \tilde{h}^{ij}(x)\tilde{h}^{kl}\rangle  
=(\delta^{ik}\delta^{jl}+\delta^{il}\delta^{jk}-\frac{2}{3}\delta^{ij}\delta^{kl})\langle \varphi(x)\varphi(x')\rangle, 
\end{align}

As discussed in the main text, the negative norm of $X$ 
%(i.e., $h^{00}$ and $\omega$) 
plays an important role 
to screen the cosmological constant. 
After setting up the problem,
we explain  the IR renormalization problem of Einstein gravity 
from the duality point of view. 
Although we can work in any conformal frame, 
we pick the following frame where the background $a$ is the classical solution. 
Sometimes we find it convenient to assume $a$ is slightly off-shell, 
\begin{align}
\frac{1}{\kappa^2}\int d^4x \big[&a^2\phi^2\tilde{R}
-6a\phi\partial_\mu\big\{\tilde{g}^{\mu\nu}\partial_\nu(a\phi)\big\}
-6H^2a^4\phi^4\big], 
\label{CFR}\end{align}
where we parametrize $\phi=e^{\omega}$. 
The quantum equation is the condition that there is no tad pole. 
In our case, it is equivalent to require that the coefficients)) in front of $\omega,\ h^{00}$ must vanish. 
In other words, there should be no linear term in $\omega,\ h^{00}$ in the effective action. 
Since de Sitter solution satisfies this requirement, 
the classical action corresponds to $\omega=h^{00}=0$, 
\begin{align}
\frac{1}{\kappa^2}\int d^4x \big[\sqrt{-\hat{g}}(\hat{R}-6H^2)
=6a\partial_0^2a-6H^2 a^4\big]. 
\label{ClAct1}\end{align}
 
There may be a gauge and parametrization where IR logarithmic effects are suppressed 
by derivative interactions. 
We perform partial integrations a few times to find such a condition. 
We suppress the $\partial Z\partial Z$-type term ($Z$ denotes $h^{\mu\nu}$ or $\omega$) in what follows. 
Such candidates are listed below 
\begin{align}
\frac{1}{\kappa^2}\int  d^4x \big[
2\partial_0a^2\tilde{g}^{0\nu}\partial_\nu\phi^2
+(6\partial_0a\partial_0a-\partial_0^2a^2)\tilde{g}^{00}\phi^2
-6H^2a^4\phi^4 \big], 
\label{DIR4}\end{align}
\begin{align}
\frac{1}{\kappa^2}\int d^4x \big[
-2\partial_0a^2 \partial_\nu\tilde{g}^{0\nu}\phi^2-a^4\hat{R}\tilde{g}^{00}\phi^2
-6H^2a^4\phi^4 \big]. 
\label{DIR5}\end{align}
The former (\ref{DIR4}) shows the equation of motion with respect to $h^{00}$ 
and the equation of motion with respect to $\phi$ is manifest in the latter (\ref{DIR5}) respectively. 

After these preparations, we integrate the IR fluctuations around the classical solution. 
The quantum equation at the one-loop level requires that no field comes out of the loop. 
So we need three-point vertices. 
The gauge fixing term is necessary only to define propagators. 
We use the exponential parametrization of the metric and quadratic gauge fixing term. 
After diagonalization, we have a massless minimally coupled mode $X$ and a conformally coupled mode $Y$. 
The latter does not have the large IR fluctuation unlike the former. 
We regard it to be constant sitting at the minimum of the potential. 
The other modes do not contribute IR logarithms to the cosmological constant. 
We decompose $h^{00}$ and $2\omega$ into
\begin{align}
h^{00}=AX+3BY,\hspace{1em}2\omega= AX+BY,\hspace{1em}
(A,B)=(\frac{\sqrt{3}}{2},\frac{1}{2}). 
\label{Sfdc}\end{align}
We need to calculate the one-point function of $\omega,\ h^{00}$ 
or take a derivative of the effective action with respect to $\omega,\ h^{00}$.
Since we are interested in IR logarithms, we can identify $h^{00}=2\omega$ for internal loop. 

We focus on a singly differentiated term in (\ref{DIR5}): 
\begin{align}
&-2\partial_0a^2\partial_\nu\tilde{g}^{0\nu}\phi^2
=2\partial_0a^2\partial_0 e^{h^{00}} e^{2\omega} \notag\\
=&2\partial_0 a^2\partial_0 e^{(AX+3BY)} e^{(AX+BY)}
=\partial_0a^2\partial_0e^{2AX+4BY}, 
\end{align}
where we assume $Y$ is constant. 
Therefore, we obtain 
\begin{align}
\int d^4x \big[\partial_0a^2\partial_0 e^{2AX}e^{4BY}
=-\partial_0^2a^2 e^{2AX}e^{4BY}
=-\partial_0^2a^2 e^{h^{00}}\phi^2\big]. 
\end{align}

We adopt the approximation $h^{00}=2\omega$ which is valid in the subspace 
with massless minimally coupled modes and the gauge we have chosen. 
It is because $h^{00}$ and $2\omega$ can be identified with a massless minimally coupled scalar $X$.  Nevertheless the original composition of the operators should enable us
to identify them as $\sqrt{-g}R$ or $\sqrt{-g}$. 

In this way, we obtain the interaction potential: 
\begin{align}
\Big[\frac{1}{\kappa^2}\int  d^4x &\big\{a^4\hat{R}-(6\partial_0a\partial_0a-\partial_0^2a^2)\big\}e^{h^{00}}\phi^2
-6H^2a^4\phi^4 \notag\\
&= \frac{1}{\kappa^2} \int  d^4x \big\{a^4\hat{R}-(6\partial_0a\partial_0a-\partial_0^2a^2)\big\}e^{2AX}e^{4BY}
-6H^2a^4e^{2AX}e^{2BY} \Big]. 
\label{DIR6}\end{align}
We can perform the same approximation in (\ref{DIR4}) with the identical result as the above. 
Note that potential vanishes on-shell in $X$ space. 
By evaluating the expectation value of the two-point functions of the interaction potential, 
we reproduce the result  (\ref{AB1}) to the linear order 
 in Appendix C.
\begin{align}
 \frac{1}{\kappa^2} \int  d^4x a^4[\hat{R}
-6He^{-2\gamma f}
-2\gamma )\partial_{\mu}f
\partial^{\mu}f]
\end{align}

We briefly recall our renormalization prescription.
In order to prepare the counter term, 
we perform the conformal transformation $a_c\rightarrow a_c^{1+3g/2}$. 
%where $g=\frac{1}{4}\frac{\kappa^2H^2}{4\pi^2}$.
We introduce an inflaton $f$ and its potential $\exp(-3g/2 f)$

to subtract noncovariant IR logarithms by a covariant counter term. 
The inflaton potential is chosen to let the classical solutions of the conformal mode and inflaton coincide. 
Since it equals $H^2a_c^{-3g/2}$, it undoes the half of the conformal transformation 
of $H^2a_c^4\rightarrow H^2a_c^{4(1+3g/2)}$. 
The remaining  overall $a_c^{3g}$ factor acts as the counter term for $\kappa^2$. 
By combining them, we reproduce the one-loop effective action $V=H^2a^{-3g \log a_c}$ 
and the solutions in the background gauge :
\begin{align}
H^2(t)&=H^2\big(1-\frac{3}{4}\frac{\kappa^2H^2}{4\pi^2}\log a_c\big), \notag\\n
a^2&=a_c^2\big(1+\frac{3}{4}\frac{\kappa^2H^2}{4\pi^2}\log a_c\big), \notag\\
\kappa^2&= \text{const}. 
\label{Qeqsol}\end{align}
These results imply the essence of our approximation is to restrict field space to that of $X$ field.

\section{Renormalization of distribution function}

\setcounter{equation}{0}

We prove the cancellations of the UV divergences and the associated time dependence of the distribution function here since it is technical.
\cite{Kitamoto2019-1}.
We assume $\rho$ can be expanded by even polynomials in $\omega$.
It is the case if $\rho$ is Gaussian.
We consider the distribution function $\rho(\omega)$.
We use dimensional regularization in $D=4-\epsilon$ dimensions.
We assume the leading quadratic divergences are absent.

Lemma:
For generic even polynomials $\rho=\omega^{2l}$,
$\rho_B= \exp\big(\tilde{g} (\frac{1}{2\epsilon}-\frac{Ht}{2})\frac{\partial^2}{\partial\omega_c^2}\big)\rho(\omega)$
 is finite and renormalization time independent:

The proof goes as follows for the fully contracted part.
\begin{align}	
&
\exp\big(\tilde{g} (\frac{1}{2\epsilon}-\frac{Ht}{2})\frac{\partial^2}{\partial\omega_c^2}\big)
\times{1\over (2l)!}<(\omega_c+\omega)^{2l}>\notag\\
=&\sum {1\over (m)!}(\tilde{g}(t) (\frac{1}{2\epsilon}-\frac{Ht}{2})^{m} {1\over (n)!}<{1\over 2}\omega^2>^{n} 
\notag\\
=&\sum {1\over (m)!}(\tilde{g}(t) (\frac{1}{2\epsilon}-\frac{Ht}{2} )^{m} {1\over (n)!}
(\tilde{g}(t) (\frac{1}{2\epsilon}-\frac{Ht}{2} ) )^{n} 
\notag\\
=& {1\over l!}(\tilde{g}(t) (\frac{1}{2\epsilon}-\frac{Ht}{2}) +\tilde{g}(t) (\frac{1}{2\epsilon}-\frac{Ht}{2}))^{l}
 \sim 0
 \label{tcan}
 \end{align}
 The coefficient for each term is uniquely determined by symmetry.
 Here $l=m+n$ and we sum over all possible $(m,n)$ and $\tilde{g}=(3g/4)$
 for simplicity. We take $\omega_c \rightarrow 0$ limit when counter terms
 are fixed. 
Since we can show quantum correction to $\rho_B $   cancel with each number of contractions in analogous way,
we conclude $\rho_B $ is finite and exit time independent.
We remark here that the above demonstrated cancelation works even if $g(t)$ is time dependent.

We may illustrate the proof inductively .
We start with the trivial identity
\begin{align}
1_B=
\exp\big(\tilde{g} (\frac{1}{2\epsilon}-\frac{Ht}{2} ) \frac{\partial^2}{\partial\omega^2}\big)\times 1 = 1
\end{align} 
Secondly
\begin{align}
(\omega)^2_B/2=
&\exp\big(\tilde{g} (\frac{1}{2\epsilon}-\frac{Ht}{2} ) \frac{\partial^2}{\partial\omega^2}\big)\times (\omega)^2/2  \notag\\
&=\omega^2/2-\tilde{g}  (\frac{1}{2\epsilon}-\frac{Ht}{2} ) \notag\\
&\rightarrow
\omega^2/2-\tilde{g} (\frac{1}{2\epsilon}-\frac{Ht}{2}) 
 +\tilde{g} (\frac{1}{2\epsilon}-\frac{Ht}{2} )
 \end{align}
In the third step
\begin{align}
 (\omega)^4
 _B/4!=
&\exp\big(\tilde{g} (\frac{1}{2\epsilon}-\frac{Ht}{2} ) )\frac{\partial^2}{\partial\omega^2}\big)\times (\omega)^4/4!
\notag\\
&= 
\omega^4/4!+\tilde{g} (\frac{1}{2\epsilon}-\frac{Ht}{2})\omega^2 /2
+(\tilde{g}(\frac{1}{2\epsilon}-\frac{Ht}{2} )^{2}/2
\notag\\
&\rightarrow
\omega^4/4! +\tilde{g}(\frac{1}{2\epsilon}-\frac{Ht}{2})\omega^2/2
+\tilde{g} (\frac{1}{2\epsilon} -\frac{Ht}{2} )\frac{\omega^2}{2}\notag\\
&
+\big(\tilde{g}(\frac{1}{2\epsilon}-\frac{Ht}{2} )
+\tilde{g}(\frac{1}{2\epsilon}-\frac{Ht}{2} )\big)^2/2
\end{align}

In each example, the second term is the subtracted operator and the last term includes the full quantum corrections.
%We make it clear how to go one more step.
$\epsilon$ and the renormalization scale (exit time) precisely cancel just as the lemma claims.
%The renormalization scale (exit time) of the low energy effective theory is the %Hubble scale.

Collorally:
We derive FP  (diffusion) equation as follows.
We  have  a prescription to renormalize the operator $\rho_B$:
\begin{align}
\rho_B	=
\exp\big(\tilde{g} (\frac{1}{2\epsilon}-\frac{Ht}{2} ) \frac{\partial^2}{\partial\omega^2}\big)\rho (\omega)
\label{rhob}
\end{align}
According the lemma, $\rho_B$ is finite and $t$ independent.
Therefore $\rho_B$ is renormalization scale (time) independent .

\begin{align}
\dot{\rho_B}= \exp\big(\tilde{g} (\frac{1}{2\epsilon}-\frac{Ht}{2}\frac{\partial^2}{\partial\omega^2}\big)
(\dot{\rho}-\frac{H}{2} \frac{\partial^2} {\partial\omega^2}\rho)
\end{align}

We conclude FP equation $(\ref{fpeq})$ holds due to renormalization scale (time)
independence. The time dependence of the gravitational coupling $\tilde{g}$
does not spoil the FP equation as the $\tilde{g}$ dependent terms: $1/\epsilon$ pole and $t$ dependence has been cancelled in (\ref{rhob})
before taking the time derivative. 
  
\begin{align}
\dot{\rho}(\omega)=\frac{H}{2} \frac{\partial^2} {\partial\omega^2}\rho
\label{fpeq}
\end{align}

\section{Scale invariant quantum gravity}
\setcounter{equation}{0}

The scale invariant theory appears at the UV fixed point.    In fact there exists such a possibility in Einstein gravity as this example indicates. We find the classical scaling theory reproduce the expected features of the quantum fixed point theory. In particular, the cosmological constant  and Higgs boson mass are protected from quantum gravity corrections.  This is a very different view
from the more conventional 'spacetime' foam.
This example demonstrates the potential significance of the duality.   

We introduce a dual pair of scaling solutions.
%This is a classical spacetime with the identical $\epsilon=3g/2$.

In the conformally flat coordinate, i.e., Poincare patch, the equations of motion are
\begin{align}
ds^2=a^2(-d\tau^2+dx_i^2), 
\end{align}
\begin{align}
a:\ \partial_0^2 a=2H^2a^3, 
\label{Eq1}\end{align}
\begin{align}
h^{00}:\ \partial_0^2 a^2=6\partial_0 a\partial_0 a. 
\label{EQMT}\end{align}

Suppose the cosmological constant evolves with time while Newton's coupling is held constant: 
\begin{align}
H^2(\tau)\propto H^2\big(\frac{1}{-H\tau}\big)^{-2\gamma}. 
\label{Hubscl}
\end{align}
Here $\gamma$ is the expansion parameter of $\Gamma,H^2,\eta$.
It is large enough to require non-perturbative analysis. 
 
According to (\ref{Hubscl}), the scale of the Universe evolves as
\begin{align}
&a=\big(\frac{1}{-H\tau}\big)^{1+\gamma}= \frac{1}{-H(\tau) \tau}, \notag\\
%&a H(\tau)=a_c H. 
\label{OmHR}\end{align}
We introduce the cosmic time $t$, 
\begin{align}
Ht= \frac{1}{\gamma}\big(\frac{1}{-H\tau}\big)^\gamma. 
\end{align}
%The scale factor is
%\begin{align}
%a=(\gamma Ht)^\frac{1+\gamma}{\gamma}. 
%\^end{align}
The Hubble parameter decays inverse proportionally with the cosmic time: 
\begin{align}
H(t)=\frac{\dot{a}}{a}=\frac{1}{\gamma t}.
%&\log a=\frac{1+\gamma}{\gamma}\log (1+\gamma Ht)\sim(1+\gamma)Ht. 
\label{Hubble1}\end{align}
The $\dot{O}$ such as $\dot{a}=\frac{\partial}{\partial t}a$ denotes 
the derivative with respect to the cosmic time $t$. 
Note that this solution doe not satisfy the other equation of motion with respect to $h^{00}$ (\ref{EQMT}) 
unless $\gamma=0$ just like 2d gravity.

This is a serious problem which needs to be addressed in order to investigate 
possible time dependence of the cosmological constant in Einstein gravity. 
Of course, such a nontrivial solution extremizes the  effective action not the tree action. 
However, Einstein action is likely to be renormalized by quantum IR effects beyond recognition. 
It may even contain new degrees of freedom.
In two dimensions, an analogous problem led us to introduce an inflaton \cite{Kitamoto2019-1} 
as a dual description of  Liouville gravity. 
A dual model is constructed in such a way that the classical evolution of an inflaton accounts for 
the quantum effects of Liouville gravity. 
We adopt the analogous strategy here and introduce an inflaton 
to satisfy the equation of motion with respect to $h^{00}$. 
Furthermore, its role is to provide a dual description of four-dimesional Einstein gravity. 
Namely we adopt the inflaton potential in such a way that the classical evolution of the inflaton reproduces 
the quantum IR effects of Einstein gravity.

As a concrete ansatz, we postulate the following Lagrangian of a single-field inflation model 
                    
\begin{align}
\frac{1}{\kappa^2}\int d^4 x\sqrt{-g}\big[{R}
-6H^2(\gamma)\exp(-2\Gamma(\gamma )f) 
-2\Gamma(\gamma)g^{\mu\nu}\partial_\mu f \partial_\nu f\big]. 
\label{AB1}
\end{align}
It is clear from this Lagrangian that the inflaton $f$ rolls down an exponential potential. 
The Hubble parameter decreases as the Universe evolves and it eventually vanishes. 
So our proposal is a de Sitter duality between quantum and classical gravitational theories. 
This action looks as follows if we make the conformal mode $a$ dependence explicit: 
\begin{align}
\frac{1}{\kappa^2}\int d^4 x\big[a^2\tilde{R} 
+6\tilde{g}^{\mu\nu}\partial_\mu a\partial_\nu a 
-6H^2(\gamma)a^4\exp(-2\Gamma(\gamma )f) 
-2\Gamma(\gamma)a^2\tilde{g}^{\mu\nu}\partial_\mu f \partial_\nu f\big], 
\label{Inflaton}\end{align}
where $H^2(\gamma)=H^2(1+\cdots)$ and $\Gamma (\gamma) = \gamma(1+\cdots)$ 
are expanded in $\gamma$

The solution is postulated to 
\begin{align}
a=e^f=a_c^{1+\gamma},~~ a_c=1/-\tau H.
\label{EQCD}\end{align}
In fact, the two coefficients, i.e., the Hubble parameter $H^2 (\gamma)$ 
and the anomalous dimension $\Gamma(\gamma) $ can be adjusted in a simple way as follows
to establish the validity of the solution (\ref{EQCD}) to  all orders in $\gamma$: 
\begin{align}
H^2(\gamma)=H^2 (1+\frac{2}{3}\gamma)(1+\gamma),\hspace{1em}
\Gamma(\gamma) = \frac{\gamma}{1+\gamma}. 
\label{anmd}
\end{align} 
The scaling dimensiom of the cosmological constant is $4-2\Gamma$.

We may sweep the inflaton under the rug by using its identity with the conformal mode (\ref{EQCD}) 
in the action (\ref{Inflaton}), 
\begin{align}
\frac{1}{\kappa^2}\int d^4 x\big[a^2\tilde{R} 
+(6-2\Gamma)\tilde{g}^{\mu\nu}\partial_\mu a \partial_\nu a 
-6H^2(\gamma)a^{4(1-\frac{\Gamma}{2})} \big]. 
\label{CMact}\end{align}
%The solution $a=a_c^{1+\gamma}$ also extremizes this restricted action 
%as it does so in an extended field space with an inflaton. 
In this Lagrangian, the nontrivial scaling dimension of the Hubble parameter 
$H^2(t)\sim a^{4-2\Gamma }$ is manifest. 
%The equation of motion with respect to $h^{00}$ is satisfied by construction. 
It requires us to introduce a new counter term. 
It is a finite renormalization of the kinetic term of the conformal mode. 
Although it is no longer manifest here, general covariance is kept intact in its dual inflation theory. 

The scaling dimension of the cosmological constant can be read off from (\ref{CMact}) as $4-2\Gamma$. This operator has the canonical scaling dimension $4$ when $\gamma=0$. As we increase $\gamma$,  it ultimately becomes 
the marginal operator with the same dimension $2$ with the Einstein term.
We note the canonical scaling also holds:$H^2(\gamma)\sim k^2 \sim \gamma^2$.
The cosmological constant cannot become irrelevant operator. 

The fixed point theory is expected to be scale invariant. We point out that our scale invariant  theory with $\Gamma=1$ is indeed a candidate for such a theory. At $\Gamma=1$  acceleration vanishes $\ddot{a}=0$.
Note that we have used the equation  of motion only to determine the dual action. Nevertheless we can extract highly nontrivial quantum effects .  
As for the Higgs boson mass, we have shown it also has the same scaling dimension $a^2$\cite{Kitamoto2014}.  It is because the masses of bosons and fermions are not renormalized by gravitational interactions at least to the one loop level. We believe there is the energy conservation law behind it. 
%If it is so, we can solve the naturalness problem of the s model of particle physics to a large extent. 
At the UV fixed point, the relative couplings of the same scaling dimensionas
 are constant under scaling.
They include $H^2/M_P^2, M_{Higgs}^2/M_P^2 $.
de Sitter duality may enable us to solve notoriously difficult quantum problems by solving much tractable Landau type classical field  theory problems with a symmetry.
%In fact we have never used quantum mechanics rections ibutionrto now. 
%We could have used quantum mechanics to evaluate scaling  c
We have derived FP equations in section 3. There is a UV complete solution.The dual pair must have the same $\epsilon$.
It is the case now with $\epsilon=\Gamma$.There are many similarities 
to the scaling solution. However $\epsilon=0$ at the fixed point of FP, while it is O(1) for the
model in this Appendix.The classical solution at the fixed point is $a=e^{\pm Ht}$.

\section{Langevin and Fokker-Planck equations}
\setcounter{equation}{0}

In this appendix, we derive FP equation from Langevin formalism.
We adopt the notation of \cite{Parisi}. 
It is straightforward to translate it to our notation.
At large times, we assume the distribution should be canonical.
\begin{align}
\dot{x}=b(t).
\end{align}
Let us consider 
\begin{align}
B(t)^{\epsilon}=\int_t^{t+\epsilon}dt' b(t').	\end{align}
We assume $\epsilon$ can be taken arbitrary small.  
In the low energy approximation, we can ignore the drift term.

Under such circumstances, 
\begin{align}
\overline{b(t_1)b(t_2)}=2\delta(t_1-t_2)A.
\label{Lancor}
\end{align}

The probability distribution of $x(t)-x(0)$ is
\begin{align}
P(x,0)=  ({1\over 4\pi tA})^{1\over 2}\exp[-{(x-x(0))^2\over 4tA}].
\end{align}
The variance is computed as
\begin{align}
\overline{(x(t)-x(0))^2}=\int_0^tdt_1\int_0^tdt_2\overline{b(t_1)b)(t_2)}
=2tA.
\end{align}

The probability $P$ satisfies the diffusion equation.
\begin{align}
	{\partial P\over \partial t}=A({\partial \over \partial x})^2P.
\end{align}
with the boundary condition
\begin{align}
P(x,0)=  \delta (x-x(0)).
\end{align}
In momentum space
\begin{align}
P(t)=e^{-\int_0^t dt' A(t') k^2}, ~~{\partial P\over \partial t}=-A(t){k}^2P.\end{align}
with the boundary condition
\begin{align}
P(k,0)= 1
\end{align}

We state a translation rule how $A$ is related to our variables 
\begin{align}
2At={3\over 4}gHt
={{3\over 4} g N} ={g\over 8\xi(t)}.
\end{align}
where $6N={1}/{\xi}$.
Thus we conclude
(\ref{FP3}) follows from the Langevin picture also.   

\section{Strong/weak coupling regime}
\setcounter{equation}{0}

The Gaussian distribution of the conformal zero mode is characterized by the standard deviation $1/\xi$.  Although there is no inflaton in Einstein gravity, 
we propose to identify the inflaton $\varphi^2$ as $\varphi^2\propto 1/\xi$. 
In our interpretation, the inflaton is not a fundamental field but a 
stochastic field. 
%parameter of the distribution function. 
It grows due to the Brownian motion:  IR logarithmic fluctuations $1/\xi\sim N $.
While the inflation theory is specified by the inflaton potential, 
the dynamics of quantum gravity is determined by the FP equation 
which describes the stochastic process at the horizon. 
We thus argue 
the classical solution of the inflation theory satisfies the FP equation as well. 

We have shown that the following linear inflaton potential is generated at the one-loop level \cite{Kitamoto2019-2} \cite{VMDT} \cite{BCFH}.
\begin{align}
\int \sqrt{-g}d^4x{1\over 16\pi G_N}
\Big[R-H^2V(\varphi)-{1\over 2}\partial_\mu \varphi\partial^\mu \varphi\Big].
\label{action2}\end{align}

Let us examine the linear potential $V(\varphi ) = 1+\sqrt{3g/2}\kappa \varphi$ due to the 1-loop quantum IR logarithm.
The slow-roll parameters are 
\begin{align}
\epsilon=(V'/V)^2/(16\pi G_N)=3g/2.
\label{slow}
\end{align}
We confirm the relation between the slow-roll parameter and 
the slope of the linear potential (\ref{slow}).
%\begin{align}
%\epsilon=\gamma
%\end{align}
Our conjecture is that the classical  inflation theory (\ref{action2}) is dual to the  IR quantum effects in Einstein gravity.
%  beyond 
%the leading one-loop IR logarithms where such a duality is observed. 
%The linear and logarithmic potentials are locally indistinguishable:  $\log(1+x)\sim x$. 

The equation of motion in the slow-roll approximation is
\begin{align}
3H(t)\dot{\varphi}=-\frac{6H^2}{\kappa}\sqrt{3g/2}. 
\label{Nleq}\end{align}
% convenience, we have take the time reversal operation here. 
Firstly, let us assume $\varphi$ is small: i.e. $\sqrt{`_3g/2}\kappa \varphi<1$.
The leading order solution is
\begin{align}
\varphi\sqrt{3g/2}\kappa=-3g Ht.
\label{Nleq}\end{align}
%The inflaton scale is safely less than $M_P$ if 
We obtain $\sqrt{4\epsilon} N=\kappa\varphi$. By squaring the both sides, we find $4\epsilon N^2 \sim N$.
In other words, $\epsilon \sim 1/4N$. 
%This is our version of the Lyth bound\cite{Lythbd}.
This estimate is consistent with the known result in the slow roll on the linear potential,  $\epsilon = 1/4N$.  

The contribution from the large inflaton field may be suppressed by small coupling $\epsilon$ since $\epsilon N$ is constant.
In the case of the convex potentials $V\sim f^m$, $\epsilon=m/4N>1/4N$ . On the other hand, $\epsilon=1/4nN<1/4N$ for concave potentials.
The former may suffer from the trans-Planckian problem while the latter may not have such a problem except $n\sim 1$.
It might be the reason why convex potentials are excluded by observations.
The dS duality is based on the possible equivalence between the slow roll in the inflation and the  random walk in the FP description. The consistency of the slow roll picture and the Brown motion picture is at the heart of dS duality. 
In the next section, we show GFP plays the key role  to bridge inflation theory
and quantum gravity at non-perturbative level.

These facts we have listed constitute the evidences for dS duality 
between quantum effects in Einstein gravity and an inflation theory (or a quintessence theory). 
It is a duality between quantum/stochastic gravity in dS space. 
%unlike dS/CFT duality \cite{Strominger2001,Maldacena2003,Balasubramanian2001}. 
It is likely that there are multiple elements in the universality class of quantum gravity/inflation theory. 
The inflation era of the early Universe may be one of them. As we discuss shortly
we find a pre-inflation era which  is indispensable to launch inflation era.
 necessary to trigger the big bang.
 Inflation theory is invented to ignite big bang in the early universe.

\end{document}